\documentclass[pra,aps,twocolumn,superscriptaddress,floatfix,noshowpacs,nofootinbib,10pt]{revtex4-2}

\usepackage[utf8]{inputenc}     	
\usepackage{amsmath, amssymb, mathtools, mathrsfs}
\usepackage[colorlinks=true]{hyperref}
\hypersetup{colorlinks,linkcolor={red},citecolor={blue},urlcolor={blue}}  
\usepackage{color}
\usepackage{graphicx}
\usepackage[normalem]{ulem}
\usepackage{mathtools}
\usepackage{stmaryrd}
\usepackage{float}
\usepackage{textalpha}

\newcommand{\bs}[1]{{\boldsymbol{#1}}}

\newcommand{\br}{\bs{r}}

\newcommand{\bq}{\bs{q}}
\newcommand{\bp}{\bs{p}}

\begin{document}
\title{Quantum kinetics of quenched two-dimensional  Bose superfluids}

\author{Cl\'ement Duval}
\email{clement.duval@lkb.upmc.fr}
\affiliation{Laboratoire Kastler Brossel, Sorbonne Universit\'{e}, CNRS, ENS-PSL Research University, 
Coll\`{e}ge de France; 4 Place Jussieu, 75005 Paris, France}

\author{Nicolas Cherroret}
\email{nicolas.cherroret@lkb.upmc.fr}
\affiliation{Laboratoire Kastler Brossel, Sorbonne Universit\'{e}, CNRS, ENS-PSL Research University, 
Coll\`{e}ge de France; 4 Place Jussieu, 75005 Paris, France}

\begin{abstract}
We study theoretically the non-equilibrium dynamics of a two-dimensional (2D) uniform Bose superfluid following a quantum quench, from its short-time (prethermal) coherent dynamics to its long-time thermalization. Using a quantum hydrodynamic description combined with a Keldysh field formalism, we derive quantum kinetic equations  for the low-energy phononic excitations of the system and characterize both their normal and anomalous momentum distributions. We apply this formalism to the interaction quench of a 2D Bose gas and study the ensuing dynamics of its quantum structure factor and coherence function, both recently measured experimentally. Our results indicate that in  two dimensions, a description in terms of independent quasi-particles becomes quickly inaccurate and should be systematically questioned when dealing with non-equilibrium scenarios.
\end{abstract}
\maketitle

\section{Introduction} 

The out-of-equilibrium dynamics of isolated quantum many-body systems has revealed a rich panel of scenarios in the recent years. In the generic case of ergodic systems, the Eigenstate Thermalization Hypothesis is expected to hold, such that at sufficiently long time any local observable acquires a value taken from a Gibbs ensemble \cite{Polkovnikov2011, Eisert2015}. In the context of experiments on cold-atomic gases, the relaxation dynamics following a quantum quench has been especially explored in one dimension, both in the weakly \cite{Gring2012, Langen2013} and strongly \cite{Kinoshita2006, Trotzky2013} interacting regimes. In the latter case, interesting theoretical predictions have also been made using non-quadratic Luttinger-liquid models \cite{Tavora2013,Protopopov2014, Buchhold2015, Buchhold2016}, such as an algebraic relaxation toward equilibrium \cite{Lin2013, Buchhold2015}. In parallel, the peculiar case of systems escaping thermalization has also attracted a lot of attention, in connection with integrability \cite{Caux2013, Bouchoule2022} or many-body localization \cite{Nandkishore2015, Alet2018, Abanin2019}.

In higher dimensions, a new generation of experiments has recently appeared, exploring, e.g.,  the relaxation dynamics of cold-atomic gases in the strong-interaction limit \cite{Eigen2017,Eigen2018} or the emergence of universal scaling laws in the vicinity of the condensation transition in three dimensions \cite{Erne2018, Glidden2021}. Concomitantly, theoretical developments  based on quantum kinetic approaches have been proposed to describe the non-equilibrium evolution  of 3D isolated quantum gases toward thermalization \cite{Griffin2009, Regemortel2018, Chantesana2019, Mikheev2019}. In comparison, on the other hand, 2D non-equilibrium Bose gases have so far received less attention. Different from 3D Bose gases, only superfluid quasi-condensates with algebraic long-range order exist for ultracold bosons in two dimensions, which requires a special treatment of phase fluctuations \cite{Popov1972, Popov1983, Mora2003}. 2D Bose gases also experience an interaction-driven Kosterlitz-Thouless transition, around which the dynamics exhibits specific temporal features \cite{Beugnon2017, Comaron2019, Sunami2022}. Generally speaking, the ability to restrict the atomic motion to two-dimensions using confining optical potentials has allowed for more and more accurate experiments of non-equilibrium physics using 2D quantum fluids \cite{Hung2013, Sunami2022, Galka2022}. 
In the context of optics, finally, a number of experiments involving ``fluids of light'' \cite{Carusotto13, Glorieux2023} have emerged, in particular in cavityless, nonlinear materials where the propagation of a laser mimics the out-of-equilibrium dynamics of 2D dilute ultracold Bose gases undergoing an interaction quench \cite{Vocke2015, Fontaine2018, Santic2018, Abuzarli2022, Steinhauer2022}.

In this paper, we present a theoretical description of the non-equilibrium quantum evolution of 2D, isolated uniform Bose superfluids following a quantum quench, which captures both the short time scales, where the dynamics is fully coherent, and the long time scales, where thermalization occurs.
To this aim, we develop a quantum kinetic formalism describing interactions between the low-lying phononic excitations of the superfluid, combining a quantum hydrodynamic representation with  a Keldysh field formalism. This allows us to go beyond recent theoretical developments based on independent quasi-particles and therefore restricted to short evolution times after the quench \cite{Natu2013, Larre2018, Martone2018, Pietraszewicz2019, Scoquart2020, Bardon-brun2020}.
Within our approach, we derive kinetic equations for both the normal and anomalous momentum distributions of the phonons, which unlike the equilibrium Bose gases are both needed to faithfully capture the non-equilibrium evolution \cite{Buchhold2015, Regemortel2018}. Close to equilibrium, in particular, we recover the Landau and Beliaev scattering rates associated with three-phonon interaction processes \cite{Beliaev1958, Pitaevskii1997, Giorgini1998, Micheli2022}. 
We finally apply this formalism to a concrete example, a quench of the interaction strength in a 2D superfluid, and analyze the subsequent time evolution of the structure factor and of the spatial coherence function, recently measured in  cold-atom \cite{Hung2013} and optical-fluid \cite{Steinhauer2022, Abuzarli2022} experiments. Our approach, in particular, 
 includes recent developments \cite{Martone2018}  allowing for a proper treatment of the finite quench duration, crucial to avoid unphysical divergences of the  post-quench superfluid's energy.

The article is organized as follows. In Sec. \ref{Sec:hydro}, we introduce the quantum hydrodynamic description of 2D superfluids and construct the non-equilibrium interacting Keldysh action in the basis of independent quasi-particles. Sec. \ref{Sec:perturbation_theory} presents the technical details on the field and perturbation theories, as well as a derivation of kinetic equations for the normal and anomalous phonon distributions. The kinetic equations and their near-equilibrium properties are discussed in Sec. \ref{Sec:kinetic}. In Sec. \ref{Sec:structure_factor}, we apply our formalism to the calculation of the the time evolution of the non-equilibrium structure factor and the coherence function of a 2D Bose gas following an interaction quench. Sec. \ref{Sec:conclusion} finally concludes the article.

\label{Sec:structure_factor}

\section{Hydrodynamic formulation}
\label{Sec:hydro}

\subsection{Hydrodynamic Hamiltonian}

Our starting point is the many-body Hamiltonian of a uniform, low-temperature, 2D gas of bosons with repulsive contact interactions,
\begin{equation}\label{microscopic_H}
	\hat{H} = \int d^2\br\Big(\!- \frac{1}{2m}  \hat{\psi}^\dagger  \Delta_{\br} \hat{\psi} + \frac{g}{2} {\hat{\psi}}^{\dagger} {\hat{\psi}}^{\dagger}  \hat{\psi} \hat{\psi}   \Big),
\end{equation}
where the field operators $\hat{\psi}$ satisfy the bosonic canonical commutation rule $[\hat{\psi}(\br),\hat{\psi}^\dagger(\br')]=\delta(\br-\br')$ and we have set $\hbar = 1$. 
In low dimension, collective excitations of the Bose gas are most conveniently described within a quantum hydrodynamic formalism, where the field operator is expressed in the density-phase representation \cite{Stringari_pitaevskii2003, Mora2003}
\begin{equation} \label{hydro_qu}
	\hat{\psi}(\br) =e^{i \hat{\theta}(\br)}\sqrt{ \hat{ \rho} (\br)} ,
\end{equation}
with the commutation rule $[ \delta \hat{ \rho} (\br), \hat{\theta} (\br')] = i \delta (\br - \br')$. At low temperature, phase fluctuations of the 2D Bose gas are generally not small, in contrast to density fluctuations and phase gradients \cite{Popov1972, Popov1983, Mora2003}.
By writing $\hat{\rho}(\br)=\rho_0+\delta\hat{\rho}(\br)$ with $\rho_0$ the mean gas density, we can then expand the Hamiltonian (\ref{microscopic_H}) with respect to $\delta\hat{\rho}$ and $\nabla_{\br} \hat{\theta}$. This leads to \cite{Popov1983, Chung2009, Bighin2015}
\begin{align} \label{hydro_Ham}
	\hat{H} = \int d \br
	 &\Big[ \frac{\rho_0}{2m}  (\nabla_{\br} \hat{\theta})^2 + \frac{g}{2} (\delta \hat{ \rho})^2 + \frac{1}{8m \rho_0} (\nabla_{\br} \delta \hat{ \rho})^2  \\
	& + \frac{1}{2m}  (\nabla_{\br} \hat{\theta}) \delta \hat{ \rho}  (\nabla_{\br} \hat{\theta}) \Big],\nonumber
\end{align}
where we have redefined the energy scale $\hat{H} \to \hat{H} - g \rho_0 / 2$ and have dropped a cubic term $\propto (\nabla_{\br} \delta \hat{ \rho})^2 \delta \hat{ \rho}$, negligible at low energy \cite{Popov1983, Chung2009}. \newline

\subsection{Bogoliubov transformation}

The Hamiltonian (\ref{hydro_Ham}) is the sum of a quadratic part $\smash{\hat{H}_0}$ and a cubic interaction term $\smash{\hat{H}_\text{int}}$. The quadratic part is non-diagonal, but is customarily diagonalized by means of a Bogoliubov transformation \cite{Altland2010}. To proceed, we first rewrite Eq. (\ref{hydro_Ham}) in momentum space, introducing the Fourier variables
\begin{equation}
	\hat{\theta}_{\bq}  \equiv \rho_0\int d \br e^{-i \bq \cdot \br }  \hat{\theta}({\br}), 
	\quad \delta\hat{\rho}_{\bq}  \equiv \int d \br e^{-i \bq \cdot \br } \hat{\rho}({\br}).
\end{equation}
The quadratic part of the Hamiltonian becomes
\begin{equation}\label{H0_Fourier}
	\hat{H}_0  =  \int_\bq\,  \Big[ \frac{\rho_0 q^2 }{2m} \hat{\theta}_{\bq} \hat{\theta}_{-\bq} + 
	 \Big( \frac{g\rho_0}{2} + \frac{q^2}{8m} \Big) { \delta \hat{\rho}}_{\bq} { \delta \hat{\rho}}_{-\bq}  \Big],
\end{equation}
where we have introduced the short-hand notation $\smash{\int_\bq\equiv \int d^2\bq/[(2\pi)^2\rho_0]}$. To diagonalize $\hat{H}_0$, we introduce new operators $\hat{a}_{\bq}$ and $\hat{a}_\bq^\dagger$, defined through the Bogoliubov transformation
\begin{align}
\label{BGtransfo1}
	{ \delta \hat{\rho}}_{\bq}^{\phantom{*}}  &= -\sqrt{\frac{E_\bq}{\epsilon_\bq}} ( \hat{a}_{\bq}^\dagger  + \hat{a}_{-\bq}^{\phantom{\dagger}} ),  \\ 
	\label{BGtransfo2}
	\hat{\theta}_{\bq}^{\phantom{\dagger}}  &= \frac{i}{2}\sqrt{\frac{\epsilon_\bq}{E_\bq}}  (  \hat{a}_{\bq}^\dagger  - \hat{a}_{-\bq}^{\phantom{\dagger}}  ),
\end{align}
where  $E_\bq\equiv \bq^2/(2m)$ and $\epsilon_\bq\equiv\sqrt{ E_\bq\left( E_\bq + 2 g \rho_0 \right)}$ is the well-known Bogoliubov dispersion relation.
Inserting this basis change into Eq. (\ref{H0_Fourier}), we obtain
\begin{equation}
\label{eq:H0}
	\hat{H}_0  =  \int_\bq\,  \epsilon_\bq \,   \hat{a}_{\bq}^\dagger \hat{a}_{\bq}^{\phantom{\dagger}},
\end{equation}
which describes a gas of free quasi-particles with energy dispersion $\epsilon_\bq$. At momenta $|\bq|\ll 1/\xi$, where $\xi\equiv \sqrt{1/4 g \rho_0 m}$ is the healing length, the dispersion relation becomes phononic: 
\begin{equation}
\label{eq_phonondispersion}
\epsilon_\bq \simeq c |\bq|,
\end{equation} 
where $c = \sqrt{g \rho_0 /m}$ is the speed of sound. Unless stated otherwise, in the rest of the paper we will mainly focus on the low-energy regime where Eq. (\ref{eq_phonondispersion}) holds. 

 In terms of the Bogoliubov operators $\hat{a}_\bq$ and $\hat{a}_\bq^\dagger$, the interaction term in the hydrodynamic Hamiltonian (\ref{hydro_Ham})  reads
\begin{equation}
\label{eq:Hint}
	\hat{H}_{\text{int}}  =  \int_{\bp,\bq}  \Lambda_{\bp, \bq}^{\phantom{*}}  \left( \hat{a}_{\bp}^{\phantom{\dagger}} \hat{a}_{\bq}^{\phantom{\dagger}} \hat{a}_{\bp+\bq}^\dagger + \text{h.c.} \right),
\end{equation}
where, in the phononic regime $|\bq|\ll1/\xi$, the vertex function $\Lambda_{\bp, \bq}$ is given by
\begin{equation}
	\Lambda_{\bp, \bq} \simeq \frac{3}{4m}  \sqrt{\frac{g \rho_0}{2 c} } \sqrt{ |\bp| \, |\bq| \, |\bp+\bq|  }.
\end{equation}
The cubic interaction (\ref{eq:Hint})  describes a three-phonon scattering process with  momentum conservation. In two dimensions it can also be resonant, i.e., there exists a range of $\bp,\bq$-values satisfying $\epsilon_\bp+\epsilon_\bq=\epsilon_{\bp+\bq}$ \cite{Andreev1980, Buchhold2015}. As will be shown in Sec. \ref{Sec:Born}, this property leads to a divergence of the self-energy, which makes this process the dominant one for the dynamics. For this reason, when writing Eq. (\ref{eq:Hint}) we have dropped interaction terms of the type $\hat{a}_{\bp}^{\phantom{\dagger}} \hat{a}_{\bq}^{\phantom{\dagger}} \hat{a}_{-\bp-\bq} $, which cannot be resonant and are therefore subdominant.

\subsection{Non-equilibrium action}

In this work, we consider a 2D Bose gas initially described by an equilibrium density matrix $\hat{\rho}_0$, and we examine its subsequent dynamics following a quantum quench performed at $t=0$. Specifically, we are interested in the time evolution of the phonon normal and anomalous momentum distributions, defined as
\begin{align}
\label{eq:nqdef}
n_{\bq,t}&\equiv \langle \hat{a}^\dagger_{\bq,t}\hat{a}_{\bq,t}\rangle\\
\label{eq:mqdef}
m_{\bq,t}&\equiv|\langle \hat{a}_{\bq,t}\hat{a}_{-\bq,t}\rangle|,
\end{align}
where the quantum-mechanical average is performed over the initial density matrix: $\langle\ldots\rangle=\text{Tr}(\hat{\rho}_0\ldots)$.

When the interaction term (\ref{eq:Hint}) in the Hamiltonian is neglected, the Heisenberg equations of motion following from Eq. (\ref{eq:H0}) lead to a purely harmonic evolution of the Bogoliubov operators, $\hat{a}_{\bq,t}=\hat{a}_{\bq,0}e^{-i \epsilon_\bq t}$, so that
\begin{equation}
n_{\bq,t}=n_{\bq,0}\quad m_{\bq,t}=m_{\bq,0}.
\end{equation}
The normal and anomalous phonon momentum distributions thus remain stuck to their initial value (more precisely, to their post-quench value, see Sec. \ref{Sec:structure_factor}), as expected from a free-field theory.

In order to capture the time dependence of $n_{\bq,t}$ and $m_{\bq,t}$ pertained to the cubic interaction (\ref{eq:Hint}), we use the Keldysh field formalism \cite{Keldysh1965, Sieberer2016}, i.e., we replace the quantum mechanical averages (\ref{eq:nqdef}) and (\ref{eq:mqdef}) by path integrals on the closed-time contour $\mathcal{C}=\{\mathcal{C}_+,\mathcal{C}_-\}$ with the forward path $\mathcal{C}_+$ ranging from $t=0$ to $\infty$ and the reversed path from $\infty$ to $0$. This amounts to doubling the degrees the freedom, i.e., we work with two sets  of scalar fields $a_+^{\phantom{*}},a_+^*$ and $a_-,a_-^{\phantom{*}}$  and the partition function
\begin{equation}
\mathcal{Z}=\int \mathcal{D}[a_+^{\phantom{*}},a_+^*,a_-^{\phantom{*}},a_-^*]e^{i S(a_+^{\phantom{*}},a_+^*)-iS(a_-^{\phantom{*}},a_-^*)},
\end{equation}
where the hydrodynamic action  in the coherent-state representation follows from Eqs. (\ref{eq:H0}) and (\ref{eq:Hint}):
\begin{align}
	&S(a,a^*) =S_0+S_\text{int}=\\
	&\int_{\bq, t}\! a_{\bq,t}^*( i \partial_t - \epsilon_{\bq} ) a_{\bq,t}^{\phantom{*}} \!+\!
	  \int_{\bp,\bq, t}\!\!\!\!  \Lambda_{\bp, \bq}^{\phantom{*}}  \left( a_{\bp,t}^{\phantom{*}} a_{\bq,t}^{\phantom{*}} a_{\bp+\bq,t}^* \!+\! \text{c.c.} \right),\nonumber
\end{align}
with the shorthand notation $\int_t=\int_{\mathcal{C}_\pm}dt$ for $a=a_\pm$. Both time integrals over $\mathcal{C}_+$ and $\mathcal{C}_-$ are conveniently reduced to a single integral over $t>0$ by introducing the ``classical'' and ``quantum'' field variables $\alpha=(a_++a_-)/\sqrt{2}$ and $\tilde\alpha=(a_+-a_-)/\sqrt{2}$ \cite{Altland2010, Kamenev2011}. 
Under this transformation, the quadratic action becomes
\begin{equation} 
\label{quadr_action_keldysh}
S_0\! = \!  \int_{\bq, t >0} \!
\begin{pmatrix}
 	\alpha^*_{\bq,t} & \tilde{\alpha}^*_{\bq,t}
 \end{pmatrix}\!
[  \bold{G}^{0}]^{-1}_{\bq,t,t}
\begin{pmatrix}
	\alpha_{\bq,t} \\
	\tilde{\alpha}_{\bq,t}
\end{pmatrix}
\end{equation}
where
\begin{equation}
\label{eq:matrixkernel}
[  \bold{G}^{0}]^{-1}_{\bq,t,t}=
\begin{pmatrix}
	0 & i \partial_t \!-\! \epsilon_\bq\! -\! i 0^+ \\
	i \partial_t \!-\! \epsilon_\bq \!+\! i 0^+ & 2i 0^+  (2 n_{\bq,0}\! +\!1)
\end{pmatrix},
\end{equation}
while the interaction part is expressed as
\begin{align}
\label{eq:SintK}
	S_{\text{int}} =& \frac{1}{\sqrt{2}} \int_{\bp,\bq,t>0}  \Lambda_{\bp, \bq}^{\phantom{*}} 
	 ( 2  \alpha_{\bp+\bq,t}^* \tilde{\alpha}_{\bp,t}^{\phantom{*}}  \alpha_{\bq,t}^{\phantom{*}} 
	  \nonumber\\
	&  + \tilde{\alpha}_{\bp+\bq,t}^* \alpha_{\bp,t}^{\phantom{*}}  \alpha_{\bq,t}^{\phantom{*}} +  \tilde{\alpha}_{\bp+\bq,t}^* \tilde{\alpha}_{\bp,t}^{\phantom{*}}  \tilde{\alpha}_{\bq,t}^{\phantom{*}}   
	  + \text{c.c.} ).
\end{align}
The Keldysh actions (\ref{quadr_action_keldysh}) and (\ref{eq:SintK}) constitute the starting point of the nonequilibrium perturbation theory  that is presented in the next section.

\section{Pertubation theory}
\label{Sec:perturbation_theory}

\subsection{Quantum kinetic equation}

To construct the perturbation theory, we introduce three fundamental correlators, the retarded $G^R$, advanced $G^A$, and Keldysh $G^K$ Green's functions: 
\begin{align}
&G^{R}_{\bq, t, t'} \equiv -i\Theta(t-t')\langle[\hat{a}_{\bq,t}^{\phantom{\dagger}},\hat{a}^\dagger_{\bq,t'}]\rangle
= -i \langle \alpha_{\bq, t}^{\phantom{*}} \tilde{\alpha}_{\bq, t'}^*  \rangle, \label{eq:GRdef}\\
&G^{A}_{\bq, t, t'} \equiv i\Theta(t'-t)\langle[\hat{a}_{\bq,t}^{\phantom{\dagger}},\hat{a}^\dagger_{\bq,t'}]\rangle
= -i \langle \tilde\alpha_{\bq, t}^{\phantom{*}} {\alpha}_{\bq, t'}^*   \label{eq:GAdef}\rangle, \\
&G^{K}_{\bq, t, t'} \equiv -i\langle\{\hat{a}_{\bq,t}^{\phantom{\dagger}},\hat{a}^\dagger_{\bq,t'}\}\rangle
= -i \langle \alpha_{\bq, t}^{\phantom{*}} {\alpha}_{\bq, t'}^*  \rangle \label{eq:GKdef}.
\end{align}
While $G^R$ and $G^A$ correspond to response functions to an external excitation, the Keldysh Green's function contains information on the system's correlations. In particular, it gives  access to the quasi-particle momentum distribution via the relation
\begin{align}
\label{eq:GK_nqt}
iG^{K}_{\bq, t, t}=2n_{\bq,t}+1,
\end{align}
deduced from Eq. (\ref{eq:nqdef}). 
The description of the anomalous distribution $m_{\bq,t}$ requires to introduce a corresponding anomalous Keldysh Green's function and is postponed to Sec. \ref{Sec:anomalous} for clarity.

In the absence of phonon interactions, the Green's functions reduce to their bare values $G^{0,R}, G^{0,A}, G^{0,K}$ and follow from Gaussian integrations on the quadratic action (\ref{quadr_action_keldysh}). This allows us to identify the elements of the matrix kernel (\ref{eq:matrixkernel}) as:
\begin{equation}
\label{eq:G0inverse}
[  \bold{G}^{0}]^{-1}\!=\!
\begin{pmatrix}
	0 &  [G^{0,A}]^{-1}\\
	 [G^{0,R}]^{-1} & -[G^{0,R}]^{-1}\!\circ\! G^{0,K}\!\circ\! [G^{0,A}]^{-1}
\end{pmatrix}
\end{equation}
and, correspondingly,
\begin{equation}
\label{eq:G0form}
 \bold{G}^{0}=
\begin{pmatrix}
	G^{0,K}& G^{0,R}\\
	G^{0,A}& 0
\end{pmatrix}.
\end{equation}
In Eq. (\ref{eq:G0inverse}), the symbol $\circ$ denotes a convolution in the time coordinates. In momentum-time representation, the bare retarded, advanced and Keldysh Green's functions take  the explicit expressions
\begin{align}
	G^{0,R}_{\bq, t, t'} &=  -i \Theta(t-t') e^{-i \epsilon_{\bq} (t-t') }, \label{G0R_expr}\\
	G^{0,A}_{\bq, t, t'} &=  i \Theta(t'-t) e^{-i \epsilon_{\bq} (t-t') }, \label{G0A_expr}\\
	G^{0,K}_{\bq, t, t'} & = - i  (2 n_{\bq,0} +1) e^{-i \epsilon_{\bq} (t-t') }. \label{G0K_expr}
\end{align}

In the presence of phonon interactions, one rewrites the total Keldysh action in the form  
\begin{equation} 
S\! = \!  \int_{\bq, t,t'} \!
\begin{pmatrix}
 	\alpha^*_{\bq,t} & \tilde{\alpha}^*_{\bq,t}
 \end{pmatrix}\!
[  \bold{G}]^{-1}_{\bq,t,t'}
\begin{pmatrix}
	\alpha_{\bq,t'} \\
	\tilde{\alpha}_{\bq,t'}
\end{pmatrix},
\end{equation}
with the matrix kernel
\begin{align}
\label{eq:gminus1}
[  \bold{G}]^{-1}_{\bq,t,t'}
=\begin{pmatrix}
	0 & [G^{0,A}]^{-1}\!-\!\Sigma^A \\
	[G^{0,R}]^{-1} \!-\!\Sigma^R  &-\Sigma^K
\end{pmatrix}_{\bq,t,t'}.
\end{align}
This structure generalizes Eq. (\ref{eq:G0inverse}) by including finite self-energies $\Sigma^{R,A,K}$ that encapsulate the effect of interactions. The self-energies can be computed from perturbation theory with the action (\ref{eq:SintK}), a task that will be undertaken in the next section. Comparing Eq. (\ref{eq:gminus1}) with the definition of $\bold{G}$, of the same triangular form as (\ref{eq:G0form}), we infer the following Dyson equations :
\begin{align}
&[G^R]^{-1}=[G^{0,R}]^{-1}-\Sigma^R\label{DysonGR}\\
&[G^A]^{-1}=[G^{0,A}]^{-1}-\Sigma^A\label{DysonGA}\\
&G^K=G^R\circ \Sigma^K\circ G^A. \label{DysonGK}
\end{align}
Within this formalism, the computation of response and correlation functions thus essentially amounts to evaluating the self-energies $\Sigma^{R,A,K}$ at a certain level of approximation.

While retarded and avanced Green's functions are both hermitian, $(G^R)^\dagger=G^R$ and $(G^A)^\dagger=G^A$, the Keldysh Green's function is anti-hermitian, $(G^K)^\dagger=-G^K$ (with the hermitian conjugate obtained by taking the complex conjugate and reversing time indices). This allows us to parametrize  $G^K$ as 
\begin{equation}
\label{eq:GK_Fparametrization}
G^{K} = G^R\circ  F - F\circ  G^A,
\end{equation}
where the hermitian distribution function $F$ will be related to the phonon momentum distribution below.
Combining Eqs. (\ref{DysonGK}) and (\ref{eq:GK_Fparametrization}), we infer:
\begin{equation}
	\Sigma^K = F \circ [G^A]^{-1} - [ G^R]^{-1}  \circ F,
\end{equation}
which, by virtue of the Dyson equations (\ref{DysonGR}) and (\ref{DysonGA}), becomes
\begin{equation}
	F\! \circ\! \big[G^{0,A}\big]^{-1}\! -\big[G^{0,R}\big]^{-1}\!  \circ\! F \!=\! \Sigma^K \!-\! \left( \Sigma^R\! \circ\! F\!  -\! F\! \circ\! \Sigma^A \right).\nonumber
\end{equation}
Direct evaluation of the left-hand side leads to the following quantum kinetic equation for the distribution function in real-time representation:
\begin{equation} \label{KE_real_time}
	i (\partial_t + \partial_{t'}) F_{\bq, t, t'} \!=\! - \Sigma^K_{\bq, t, t'} \!+\! \left( \Sigma^R\! \circ\! F \!-\! F\! \circ\! \Sigma^A \right)_{\bq, t, t'}.
\end{equation}
An evaluation of this evolution equation requires the knowledge of the Keldysh and retarded self-energies, which will be both computed in Sec. \ref{Sec:Born}. Before that, we introduce an important assumption that will bring about a first important simplification of  Eq. (\ref{KE_real_time}).\newline

\subsection{Separation of time scales and on-shell approximation}
\label{eq:time_separation}

Two-time non-equilibrium functions such as $F_{\bq, t, t'}$ are most conveniently expressed using the Wigner coordinates $\tau\equiv (t+t')/2$ and $\Delta t\equiv t-t'$. The Wigner transform of a given two-time function $X_{t,t'}$ is defined as $X_{\omega, \tau} = \int d \Delta t \,  e^{i \Delta t \omega} X_{\tau + \Delta t/2, \tau - \Delta t /2}.$
In the present context, the central time $\tau$ is associated with the slow relaxation of the phonons, while the time difference $\Delta t$ is related to their fast, coherent dynamics \cite{Honeychurch2019}. 

In the presence of interactions, Bogoliubov quasi-particles acquire a finite lifetime $\tau_\bq\sim -1/\text{Im}\Sigma^R_\bq$. As long as interactions are weak, this lifetime is  typically very long compared to the coherent time scale $1/\epsilon_\bq$ :
\begin{equation}
\label{eq:separationtime}
\tau_\bq\sim -1/\text{Im}\Sigma^R_\bq\gg 1/\epsilon_\bq.
\end{equation}
This condition, which we will verify \emph{a posteriori} below, also implies that quasi-particles remain well defined during of the out-of-equilibrium evolution. In the limit (\ref{eq:separationtime}), it can be shown that the Wigner transform of a time convolution reduces [at leading order in $1/(\epsilon_\bq \tau_\bq)\ll1$] to the product of Wigner transforms \cite{Kamenev2011}. The Wigner transform of Eq. (\ref{KE_real_time}) thus simplifies to
\begin{equation}
\label{relation_GK_F}
i\partial_\tau F_{\bq, \omega,\tau} \!\simeq\! - \Sigma^K_{\bq,\omega, \tau} \!+\! 2iF_{\bq, \omega,\tau} \text{Im}(\Sigma^R_{\bq, \omega, \tau}).
\end{equation}

Within the separation of time scales (\ref{eq:separationtime}), application of the Wigner transform to Eq. (\ref{eq:GK_Fparametrization}) also yields:
\begin{align}
\label{GK_Wingerapprox}
iG^K_{\bq,\omega,\tau}\simeq F_{\bq,\tau,\omega}A_{\bq,\omega,\tau},
\end{align}
where 
\begin{align}
A_{\bq,\omega,\tau}
\equiv-2 \, \text{Im}(G^R_{\bq,\omega,\tau})
=\frac{-2  \, \text{Im}(\Sigma^R_{\bq,\omega,\tau})}{|\omega-\epsilon_\bq-\Sigma^R_{\bq,\omega,\tau}|^2}.
\end{align}
$A_{\bq,\tau,\omega}$ is the spectral function, which gives the probability density that a quasi-particle with energy $\omega$ has the dispersion $\epsilon_\bq$ at a time $\tau$ after the quench. Under the condition (\ref{eq:separationtime}) of well defined quasi-particles, the spectral function is strongly peaked around $\omega=\epsilon_\bq$ [with $A_{\bq,\tau,\omega}\to 2\pi\delta(\omega-\epsilon_\bq)$ in the non-interacting limit]. Integrating Eq. (\ref{GK_Wingerapprox}) over $\omega$ then leads to
\begin{equation}
\label{eq:GK_onshell}
\int\frac{d\omega}{2\pi}iG^K_{\bq,\omega,\tau}\simeq F_{\bq,\epsilon_\bq,\tau}=2 n_{\bq,\tau}+1,
\end{equation}
where we have used Eq. (\ref{eq:GK_nqt}) in the last equality. This relation shows that as long as the separation of time scales (\ref{eq:separationtime}) holds, the phonon momentum distribution coincides with the distribution function $F_{\bq,\omega, \tau}$ evaluated at $\omega=\epsilon_\bq$, a property known as the on-shell approximation. To evaluate $n_{\bq,\tau}$, it is thus sufficient to solve the on-shell version of the kinetic equation (\ref{KE_real_time}). This is achieved by multiplying the latter by the spectral function and integrating over $\omega$, similarly to Eq. (\ref{eq:GK_onshell}). The kinetic equation simplifies to:
\begin{equation}
\label{eq:kinetic_onshell}
\partial_\tau F_{\bq, \tau} \!\simeq\! i\Sigma^K_{\bq, \tau} \!+\! 2F_{\bq, \tau} \text{Im}(\Sigma^R_{\bq, \tau}),
\end{equation}
where we have introduced the simpler notations $F_{\bq, \tau}\equiv F_{\bq,\epsilon_\bq,\tau}$ and $\Sigma_{\bq, \tau}\equiv \Sigma_{\bq,\epsilon_\bq,\tau}$. Together with Eq. (\ref{eq:GK_onshell}), Eq. (\ref{eq:kinetic_onshell}) constitutes a quantum kinetic equation for the momentum  distribution of the interacting phonons, which can be directly solved once an approximation for the self-energies is provided.

\subsection{Born approximation}
\label{Sec:Born}

We now evaluate the retarded and Keldysh self-energies $\Sigma^R_{\bq, \tau}$ and  $\Sigma^K_{\bq, \tau}$ for a 2D, weakly interacting Bose gas using perturbation theory. In practice, this is achieved by expanding $e^{i S_{\text{int}}}$ and truncating the corresponding series at leading order. For a dilute Bose gas, $e^{i S_{\text{int}}}$ can be expanded at first order in the interaction parameter $g \rho_0$ (Born approximation), such that the retarded Green's function is approximated as
\begin{align}\label{correlator_r_Born}
	G^R_{\bq,t,t'}&\!=\!-i\langle \alpha_{\bq, t}^{\phantom{*}}  \tilde{\alpha}_{\bq, t'}^* \rangle  
	\!=\!-i \int \mathcal{D}[ \alpha , \tilde{\alpha}] \alpha_{\bq, t}^{\phantom{*}} \tilde{\alpha}_{\bq, t'}^* e^{i \left( S_0 + S_\text{int} \right)}\nonumber\\
	&\!\simeq\!-i \int \mathcal{D}[ \alpha , \tilde{\alpha}] \alpha_{\bq, t}^{\phantom{*}} \tilde{\alpha}_{\bq, t'}^* e^{i  S_0}(1-S_\text{int}^2/2).
\end{align}
Comparison with Eq. (\ref{DysonGR}) then provides an explicit expression for the self-energy.
The Gaussian integral in Eq. (\ref{correlator_r_Born}) yields three contributions to $\Sigma^R$, each appearing with multiplicity 8 and represented by the one-loop diagrams in Fig. \ref{fig_diagrams}(b) [see Fig. \ref{fig_diagrams}(a) for the diagrammatic conventions]:
\begin{figure}
\includegraphics[scale=0.92]{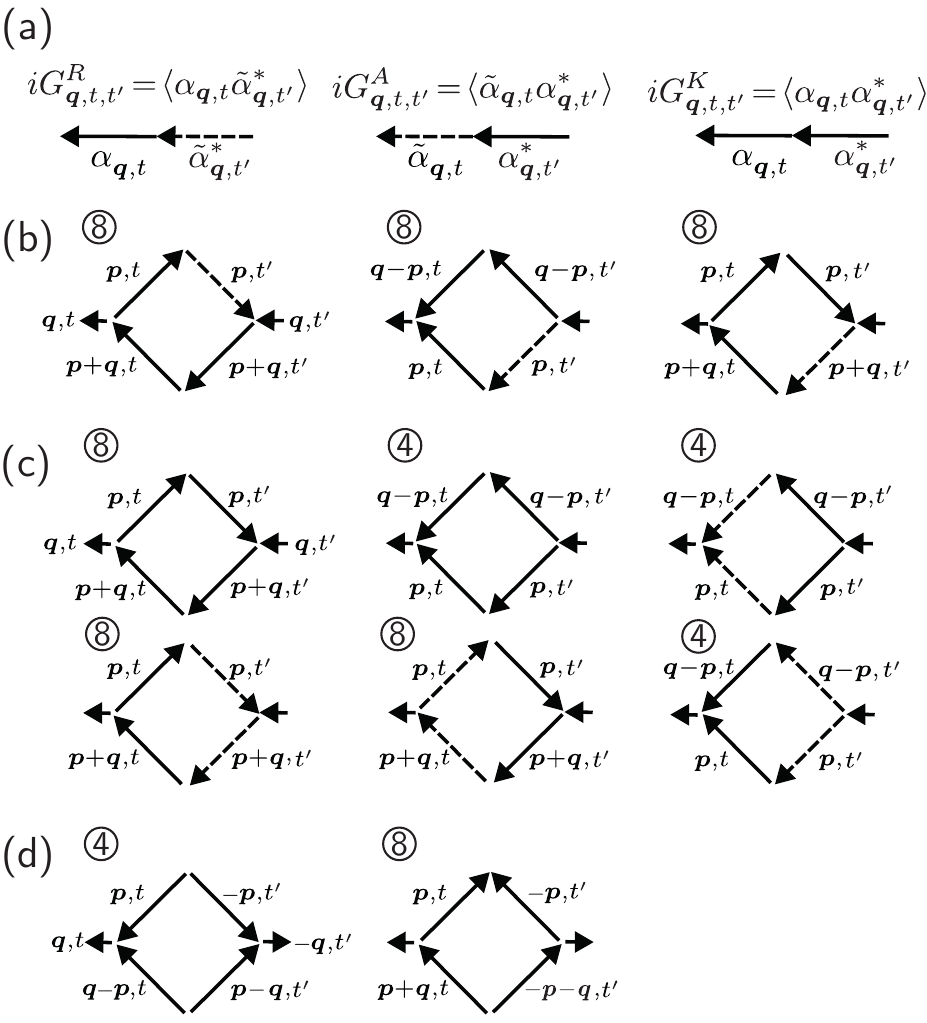}
\caption{(a) Diagrammatic conventions for the Green's functions. Dashed (solid) lines refer to a quantum $\tilde\alpha$ (classical $\alpha$) field variable.  Arrows are directed from a conjugated field variable to a non-conjugated one.
(b) Diagrams contributing to the retarded self-energy $\Sigma^R$ [Eq. (\ref{eq:sigmaR_Born})]. 
Each diagram has multiplicity 8.  (c) and (d) are the diagrams contributing to the normal $\Sigma^K$ and anomalous $\mathscr{S}^K$ Keldysh self-energies, with the corresponding multiplicities indicated. 
\label{fig_diagrams}}
\end{figure}
\begin{align}
\label{eq:sigmaR_Born}
	\Sigma^R_{\bq, t, t'} &= 2i \int_{\bp}   \Big[
	\Lambda_{\bp, \bq}^2 G^{K}_{\bp+\bq, t, t'} G^{0,A}_{ \bp, t', t} \\
	 &+ \Lambda_{\bp, \bq-\bp}^2 G^{K}_{\bq-\bp, t, t'} G^{0,R}_{\bp,t, t'} 
	+  \Lambda_{\bp, \bq}^2 G^{K}_{\bp, t',t} G^{0,R}_{\bp+ \bq, t, t'}\Big] .\nonumber
\end{align}
In the Wigner representation, this reads:
\begin{align}
\label{eq:sigmaR_Born2}
\Sigma_{\bq, \omega,\tau}^R &= 
2 i \int_{\bp, \nu}\Big\{   \Lambda^2_{\bp, \bq-\bp} G^{K}_{\bp, \nu,\tau} G^{0,R}_{\bq-\bp, \omega - \nu} \\
&+ \Lambda^2_{\bp, \bq} \left[ G^{K}_{\bp, \nu} G^{0,R}_{\bp+\bq, \omega + \nu} + G^{K}_{\bp+\bq, \nu + \omega,\tau} G^{0,A}_{\bp,  \nu} \right]\Big\}.\nonumber
\end{align}
Next, we use that $G^{0,R}_{\bp,\nu}=-i\pi \delta(\nu-\epsilon_\bp)$, $G^{0,A}_{\bp,\nu}=i\pi \delta(\nu-\epsilon_\bp)$ and $G^K_{\bp,\tau,\nu}\simeq -2i\pi F_{\bp,\nu,\tau}\delta(\nu-\epsilon_\bp)$ [cf Eqs. (\ref{G0R_expr}, \ref{G0A_expr}, \ref{G0K_expr})], multiply Eq. (\ref{eq:sigmaR_Born2}) by the spectral function $A_{\bq,\omega,\tau}$ and integrate over $\omega$ and $\nu$ using that the latter is peaked around $\omega\simeq\epsilon_\bq$. This yields the on-shell self-energy
\begin{align} 
\label{Sigma_R_after_nu_integration}
\Sigma_{\bq,\tau}^R &\simeq   -2 i \pi \int_{\bp} \Big[\Lambda^2_{\bp, \bq}
  \left(  F_{\bp,\tau}  -   F_{\bp+\bq,\tau} \right)  \delta(\epsilon_{\bq} + \epsilon_{\bp} - \epsilon_{\bp + \bq})\nonumber\\
  & + \Lambda^2_{\bp, \bq-\bp}  F_{\bp,\tau } \,   \delta(\epsilon_{\bq} - \epsilon_{\bp} - \epsilon_{\bq-\bp})\Big]. 
\end{align}
For a purely phononic dispersion (\ref{eq_phonondispersion}), the angular integration in Eq. (\ref{Sigma_R_after_nu_integration}) is divergent, which is a consequence of the resonant character of the cubic interaction (\ref{eq:Hint}). In two and three dimensions,  this divergence is customarily regularized by taking into account the first nonlinear correction to the Bogoliubov dispersion, $\epsilon_\bq\simeq c|\bq|+(c\xi^2/2)|\bq|^3$ \cite{Pitaevskii1997, Chung2009}. Note that in strongly-interacting gases in one dimension, it has been suggested that the divergence should be instead resolved via a self-consistent Born approximation \cite{Buchhold2015}. In the present case of a dilute Bose gas, however, such an approach would lead to sub-leading contributions and is therefore not adequate. 
Including the leading-order corrections to the linear dispersion and performing the angular integrations in Eq. (\ref{Sigma_R_after_nu_integration}), we finally obtain
\begin{equation}
\label{eq:sigmaR_final}
	\Sigma^R_{\bq,\tau}\! =\! -\frac{i}{2}  \int_{0}^\infty \!\!dp \,  \mathcal{ K}^L_{\bp,\bq} (F_{\bp,\tau} \!-\! F_{\bp+ \bq,\tau} )\!  -\! i \int_{0}^q \! dp \, \mathcal{ K}^B_{\bp,\bq} F_{\bq,\tau}
\end{equation}
where 
\begin{equation}
\label{eq:def_KBL}
	\mathcal{K}^L_{\bp,\bq} = \frac{3 \sqrt{3} c}{8 \pi \rho_0 } p (p+q), \quad \mathcal{K}^B_{\bp,\bq} = \frac{3 \sqrt{3} c}{16 \pi \rho_0 } p (q-p).
\end{equation}

We now come to the Keldysh self-energy $\Sigma^K$, which is calculated perturbatively from  the Dyson equation (\ref{DysonGK}). At the Born approximation, this is achieved by 
approximating the left-hand side by
\begin{align}\label{correlator_r_def}
	{G}^K_{\bq,t,t'}\!\simeq\!-i \int \mathcal{D}[ \alpha]\, \alpha_{\bq, t}^{\phantom{*}} {\alpha}^*_{\bq, t'}e^{i  S_0}(1-S_\text{int}^2/2).
\end{align}
Evaluation of the Gaussian integral involves the six one-loop diagrams represented in Fig. \ref{fig_diagrams}(c):
\begin{flalign}
	\Sigma^K_{\bq, t, t'} &\!=  \! i \int_{\bp} 
	\Big[2 \Lambda_{\bp, \bq}^2 \big(  G^{K}_{\bp+\bq, t, t'} G^{K}_{\bp, t', t}  +  G^{0,A}_{\bp, t', t} G^{0,R}_{\bp+\bq, t, t'} \nonumber\\
	&+   G^{0,R}_{\bp, t', t} G^{0,A}_{\bp+ \bq, t, t'}  \big)
	 +  \Lambda_{\bp, \bq- \bp}^2 \big(  G^{K}_{\bq-\bp, t, t'} G^{K}_{\bp, t, t'}\nonumber\\
	 &+ G^{0,A}_{\bq-\bp, t, t'} G^{0,A}_{\bp, t, t'} + G^{0,R}_{\bq-\bp, t, t'} G^{0,R}_{\bp, t, t'}  \big)\Big]. 
	 \label{eq:sigmaK_general}
\end{flalign}
To evaluate this expression, we proceed as for $\Sigma^R$, namely, we move to Wigner representation, multiply Eq. (\ref{eq:sigmaK_general}) by the spectral function and integrate over Wigner frequencies. This leads to the on-shell value
\begin{align}
	\Sigma^K_{\bq,\tau} &=  \, - 2 i \pi  \int_{\bp} \Big[ 2\Lambda_{\bp, \bq}^2  \, \left(  F_{\bp+\bq} F_{\bp } -1 \right) 
	\delta(\epsilon_{\bq} + \epsilon_{\bp} - \epsilon_{\bp+\bq}) \nonumber \\
	& +  \Lambda_{\bp, \bq- \bp}^2  \,  \left(  F_{\bq-\bp} F_{\bp } +1
	\right)   \delta(\epsilon_{\bq} - \epsilon_{\bp} - \epsilon_{\bq-\bp})\Big].
\end{align}
By finally computing  the angular integration using the regularization procedure explained above, we find:
\begin{align}
\label{eq:sigmaK_final}
	\Sigma^K_{\bq,\tau} =&  - i  \int_0^\infty dp \, \mathcal{K}^L_{\bp,\bq} \left(    F_{\bp+\bq,\tau}  F_{\bp,\tau}   -1    \right)\nonumber\\
	& - i   \int_0^q  dp \, \mathcal{K}^B_{\bp,\bq} \left(    F_{\bq-\bp,\tau}  F_{\bp,\tau}   + 1    \right)  .
\end{align}
Equations (\ref{eq:sigmaR_final}) and (\ref{eq:sigmaK_final}) constitute the final expressions for the normal self-energies, which once inserted in Eq. (\ref{eq:kinetic_onshell}) provide a kinetic equation for the momentum distribution $n_{\bq,\tau}$. Before coming to that point, however, we now discuss the perturbation theory for the anomalous distribution.

\subsection{Anomalous momentum distribution}
\label{Sec:anomalous}

In order to derive a quantum kinetic equation for the anomalous phonon distribution $m_{\bq,\tau}$, one is naturally led to define a Keldysh Green's function from the anomalous anti-commutator 
$\langle
\{\hat{a}_{\bq,t},\hat{a}_{-\bq,t'}\}\rangle$, in analogy with Eq. (\ref{eq:GKdef}). Such a definition, however, gives rise to fast temporal oscillations at the scale of $1/\epsilon_\bq$, which are incompatible with the requirement of time scales separation discussed in Sec. \ref{eq:time_separation}. This can be seen at the level of the free-field theory, which yields 
$\langle\{\hat{a}_{\bq,t},\hat{a}_{-\bq,t'}\}\rangle=2\langle\hat{a}_{\bq,0}\hat{a}_{-\bq,0}\rangle\exp(-2i\epsilon_\bq\tau)$, where $\tau=(t+t')/2$. To get rid of these fast variations, we move to the rotating time frame by employing the transformation  $\alpha_{\bq,t}\to \alpha_{\bq,t}\exp(i\epsilon_\bq t)$, following \cite{Buchhold2015, Buchhold2016}. In this rotating frame, we can safely define the anomalous Keldysh Green's function as 
\begin{equation}
i\mathscr{G}^K_{\bq,t,t'}=
\langle\{\hat{a}_{\bq,t},\hat{a}_{-\bq,t'}\}\rangle=
\langle\alpha_{\bq,t}\alpha_{-\bq,t'}\rangle.
\end{equation}
From its definition (\ref{eq:mqdef}), the anomalous momentum distribution follows from $2m_{\bq,t}=i\mathscr{G}^K_{\bq,t,t}$ \cite{footnote}. In the presence of phonon interactions, $\mathscr{G}^K$ acquires a finite (anomalous) self-energy  $\mathscr{S}^K$, defined through a Dyson equation similar to Eq. (\ref{DysonGK}):
\begin{equation}
\label{eq:anomalous_Dyson}
\mathcal{G}^K=G^R\circ\mathscr{S}^K\circ {G}^A.
\end{equation}
In the rotating frame, $\mathscr{G}^K$ is also anti-hermitian, and can therefore be parametrized in a similar way as $G^K$ [see Eq. (\ref{eq:GK_Fparametrization})]:
\begin{equation}
\label{eq:gK_Fparametrization}
\mathscr{G}^K = G^R \circ  \mathscr{F} - \mathscr{F}   \circ G^A.
\end{equation}
Combining the four relations (\ref{DysonGR}), (\ref{DysonGA}), (\ref{eq:anomalous_Dyson}) and (\ref{eq:gK_Fparametrization}) and making use of the condition of time scales separation, as explained in Sec. \ref{eq:time_separation}, we infer the anomalous version of the on-shell kinetic equation (\ref{eq:kinetic_onshell}):
\begin{equation}
\label{eq:kinetic_anomalous_onshell}
\partial_\tau \mathscr{F}_{\bq, \tau} \!\simeq\! i\mathscr{S}^K_{\bq, \tau} \!+\! 2\mathscr{F}_{\bq, \tau} \text{Im}(\Sigma^R_{\bq, \tau}).
\end{equation}
As compared to Eq. (\ref{eq:kinetic_onshell}), the difference lies in the  anomalous Keldysh  self-energy $\mathscr{S}^K_{\bq, \tau}$, which can be computed by perturbation theory from Eq. (\ref{eq:anomalous_Dyson}). Similarly to the normal correlator, this is done by approximating the left-hand side of Eq. (\ref{eq:anomalous_Dyson}) by
\begin{align}\label{correlator_r_def2}
	\mathcal{G}^K_{\bq,t,t'}\!\simeq\!-i \int \mathcal{D}[ \alpha]\, \alpha_{\bq, t} {\alpha}_{-\bq, t'}e^{i  S_0}(1-S_\text{int}^2/2).
\end{align}
Note that different from the calculation of $\Sigma^K$, here the Wick decomposition following from the Gaussian integral (\ref{correlator_r_def2}) is performed by only considering pairings of anomalous correlators, which is a consequence of the resonant character of the interaction, see \cite{Buchhold2015} for details. This leads to the two self-energy diagrams displayed in Fig. \ref{fig_diagrams}(d), which explicitly read: 
\begin{align}
	\mathscr{S}^K_{\bq, t, t'}\! =\!i\! \int_{\bp} [2 \Lambda_{\bp, \bq}^2   \, \mathscr{G}^{K}_{\bp+\bq, t, t'} \mathscr{G}^{K}_{\bp, t', t}
	 \!+\!\Lambda_{\bp, \bq- \bp}^2 \,  \mathscr{G}^{K}_{\bp, t, t'}  \mathscr{G}^{K}_{\bq-\bp, t, t'} ].\nonumber 
\end{align}
We then proceed as in Sec. \ref{Sec:Born}, i.e., we compute the Wigner transform of $\mathscr{S}^K_{\bq, t, t'}$, multiply by the spectral function and integrate over Wigner frequencies. It eventually yields
\begin{align}
\label{eq:sigmaKanomalous_final}
	\mathscr{S}^K_{\bq,\tau}=& - i\int_0^\infty \!\!dp \mathcal{K}^L_{\bp,\bq}    \mathscr{F}_{\bp+\bq,\tau}  \mathscr{F}_{\bp,\tau}    \nonumber  \\
	&-i  \int_0^q  \!dp \, \mathcal{K}^B_{\bp,\bq}   \mathscr{F}_{\bq-\bp,\tau}  \mathscr{F}_{\bp,\tau}.
\end{align}
Once inserted in (\ref{eq:kinetic_anomalous_onshell}), this provides an explicit expression for the kinetic equation for $m_{\bq,\tau}$.

\section{Phonon quantum kinetics}
\label{Sec:kinetic}

\subsection{Kinetic equations}

Inserting the expressions (\ref{eq:sigmaR_final}) and (\ref{eq:sigmaK_final}) of the normal self-energies into Eq. (\ref{eq:kinetic_onshell}) and using Eq. (\ref{eq:GK_onshell}), we obtain the final form of the kinetic equation for the phonon momentum distribution $n_{\bq,\tau}$ at the Born approximation:
\begin{align}\label{KE_born_nq}
	\partial_{\tau} n_{\bq,\tau}& =  2 \int_0^\infty dp \, \mathcal{K}^L_{\bp,\bq} \left[  n_{\bp+\bq} \left( n_\bp \!+\! n_\bq \!+\!1  \right)\! -\! n_\bp n_\bq  \right] +\nonumber\\
	& 2 \int_0^q  dp \, \mathcal{K}^B_{\bp,\bq} \left[ n_\bp n_{\bq-\bp} \!-\! n_\bq \left( n_\bp \!+\! n_{\bq-\bp} \!+\!1 \right) \right],
\end{align}
where for notation simplicity we have dropped the $\tau$ indices in the collision integrals, and we recall that the  kernels $\mathcal{K}^L_{\bp,\bq}$ and $\mathcal{K}^B_{\bp,\bq}$ are  given by Eq. (\ref{eq:def_KBL}). 
The only quantity conserved during the time evolution pertained to Eq. (\ref{KE_born_nq}) is the (phononic) energy: $\partial_\tau\int_\bq c|\bq| n_{\bq,\tau}=0$ for all $\tau$. The kinetic equation includes two collision integrals, which correspond to the well-known Beliaev $(B)$ and Landau $(L)$ three-particle scattering processes. The Beliaev process  describes the splitting $q\to(p,q-p)$ of the probe phonon of momentum $q$ into two phonons of momenta $p$ and $q-p$, while the Landau process describes the recombination $(q,p)\to p+q$ of the probe phonon with another one, each process coming together with its reversed version. Both Landau and Beliaev processes lead to a relaxation of the momentum distribution toward a thermal equilibrium at long time:
\begin{equation}
\label{eq:nq_thermal}
n_{\bq,\tau\to\infty}\equiv n_{\bq}^\text{th}=\frac{1}{\exp(\epsilon_\bq/T)-1},
\end{equation}
a solution which cancels both collision integrals in Eq. (\ref{KE_born_nq}). 
In the present non-equilibrium scenario, the temperature $T$ characterizes the effective thermal equilibrium reached by the phonon gas a long time after the quench. In practice, this temperature is determined from the law of energy conservation mentioned above. A concrete example of this will be given  in Sec. \ref{Sec:structure_factor}.

The kinetic equation for the anomalous phonon distribution is similarly derived, by inserting Eq. (\ref{eq:sigmaR_final}) and (\ref{eq:sigmaKanomalous_final}) into Eq. (\ref{eq:kinetic_anomalous_onshell}):
\begin{align}\label{KE_born_mq}
	\partial_{\tau} m_{\bq,\tau}& =  2 \int_0^\infty dp \, \mathcal{K}^L_{\bp,\bq} 
	\left(  n_{\bp+\bq}m_\bq\! +\! m_\bp m_{\bp+\bq}-n_\bp m_\bq  \right) 
	+\nonumber\\
	& 2 \int_0^q  dp \, \mathcal{K}^B_{\bp,\bq} \left[ m_\bp m_{\bq-\bp} \!-\! m_\bq \left( n_\bp \!+\! n_{\bq-\bp} \!+\!1 \right) \right].
\end{align}
Notice that the dynamics of the anomalous distribution is coupled to the evolution of $n_{\bq,\tau}$. Furthermore, unlike $n_{\bq,\tau}$ the anomalous distribution vanishes at long time:
\begin{equation}
\label{eq:mq_thermal}
m_{\bq,\tau\to\infty}=0,
\end{equation}
a result expected for a quantum gas at thermal equilibrium. At any finite time, however, $m_{\bq,\tau}$ is in general nonzero and may significantly impact the intermediate-time dynamics of  physical observables.

\subsection{Near-equilibrium solutions}

Before examining the consequences of the phonon relaxation on a concrete scenario, it is useful to discuss the near-equilibrium case, which for a quench experiment typically corresponds to the long-time regime where the distributions $m_{\bq,\tau}$ and $n_{\bq,\tau}$ become close to their equilibrium values (\ref{eq:nq_thermal}) and $(\ref{eq:mq_thermal})$. 
To this aim, we substitute $n_{\bq,\tau}=n_\bq^\text{th}+\delta n_{\bq,\tau}$ with $\delta n_{\bq,\tau}\ll n_\bq^\text{th}$ in the kinetic equation (\ref{KE_born_nq}) and linearize. If only the Beliaev collision integral is kept, this leads to 
\begin{equation}
\label{eq:gammaB}
\partial_\tau \delta n_{\bq,\tau}\simeq-2\gamma^B_\bq \delta n_{\bq,\tau},\ \ \gamma^B_\bq=\frac{\sqrt{3}c}{32\pi\rho_0}q^3.
\end{equation} 
This describes an exponential relaxation governed by the Beliaev damping rate $\gamma_\bq^B$. Note that, alternatively, the latter could have been directly derived from the self-energy (\ref{eq:sigmaR_final}) evaluated at equilibrium: $\smash{\gamma_\bq^B=-\text{Im}\Sigma^R_\bq(n_\bq^\text{th})}$. 
In two dimensions, the Beliaev damping rate (\ref{eq:gammaB}) has been previously derived in \cite{Chung2009} using the Matsubara formalism. 

If, on the other hand, only the Landau collision integral in Eq. (\ref{KE_born_nq}) is considered, the linearization procedure provides
\begin{equation}
\label{eq:gammaL}
\partial_\tau \delta n_{\bq,\tau}\simeq-2\gamma^L_\bq \delta n_{\bq,\tau},\ \ \gamma^L_\bq=\frac{\sqrt{3}\pi}{8\rho_0 c}q T^2,
\end{equation} 
which now involves the Landau damping rate $\gamma^L_\bq$ \cite{Chung2009}. Comparison of Eqs. (\ref{eq:gammaB}) and (\ref{eq:gammaL}) shows that Beliaev processes are mostly effective when the long-time equilibrium temperature vanishes. Landau processes, on the contrary, typically dominate at finite temperature. In the relaxation following a quantum quench, this is the most common situation since the quench inevitably injects a certain amount of energy into the system, eventually leading to a finite-temperature state at long time.

A near-equilibrium expansion, finally, can also be used for the anomalous momentum distribution, using that $m_{\bq,\tau}\ll1$ at long time. Expanding Eq. (\ref{KE_born_mq}) for small $\delta n_{\bq,\tau}$ and small $m_{\bq,\tau}$ then yields 
\begin{equation}
\label{eq:gammaL_mq}
\partial_\tau  m_{\bq,\tau}\simeq -2\gamma^{L,B}_\bq  m_{\bq,\tau},
\end{equation} 
depending on which of the Beliaev or Landau processes dominates.
\newline

\section{Application: non-equilibrium structure factor and coherence}
\label{Sec:structure_factor}

\subsection{Quench protocol}

We now apply the above formalism to a concrete scenario. Consider a uniform two-dimensional Bose gas, initially at equilibrium at temperature $T_0$ in a superfluid state  with interaction strength $g_0$. The initial (pre-quench) quasi-particle distributions are thus given by
\begin{equation}
\label{eq:nq0_BE}
n_{\bq}^0=\frac{1}{\exp(\epsilon_\bq^0/T_0)-1},\ \ m_{\bq}^0=0,
\end{equation}
where $\epsilon_\bq^0=\sqrt{E_\bq(E_\bq +2g_0\rho_0)}$, with $E_\bq=\bq^2/2m$. As a quench protocol, we suppose that around the time $\tau=0$ the interaction strength is changed from $g_0>0$ to another positive value $g\ne g_0$. The simplest description of this problem, studied, e.g., in \cite{Rancon2013, Martone2018},  consists in assuming that the interaction change occurs \emph{instantaneously} at $\tau=0$. In that case, applying the Bogoliubov transformation (\ref{BGtransfo1}) at both $\tau=0^-$ and $\tau=0^+$ and using the continuity of the field operator (\ref{hydro_qu}), we obtain the following relation between the post-quench ($\hat{a}_{\bq}^{\text{ps} } ,\hat{a}_{\bq}^{\text{ps}\dagger } $) and pre-quench ($\hat{a}_{\bq}^0  ,\hat{a}_{\bq}^{0\dagger } $) Bogoliubov operators:
\begin{equation}
	\begin{pmatrix}
		\hat{a}_{\bq}^{\text{ps}\dagger } \vspace{0.1cm}\\
		\hat{a}_{-\bq}^{\text{ps} \phantom{\dagger}}
	\end{pmatrix}
	= \dfrac{1}{2 \sqrt{\cramped \epsilon_\bq^{\phantom{0}}  \epsilon_\bq^0 } }
	\begin{pmatrix}
		\epsilon_{\bq}^{\phantom{0}}  + \epsilon_\bq^0  &  \epsilon_{\bq}^{\phantom{0}}  - \epsilon_\bq^0   \vspace{0.1cm}\\
		\epsilon_{\bq}^{\phantom{0}}  - \epsilon_\bq^0 &  	\epsilon_{\bq}^{\phantom{0}}  + \epsilon_\bq^0
	\end{pmatrix}
	\begin{pmatrix}
		\hat{a}_{\bq}^{0\dagger  }  \vspace{0.1cm}\\
		\hat{a}_{-\bq}^{0\phantom{\dagger}}\nonumber
	\end{pmatrix}.
\end{equation}
where $\epsilon_\bq=\sqrt{E_\bq(E_\bq+ 2g\rho_0)}$.
The post-quench normal and anomalous momentum distributions then take the form
\begin{align} 
\label{smooth_quench_nq}
	n_{\bq}^\text{ps}& = n_{\bq}^0 \left(2 d_\bq^2 +1 \right) + d_\bq^2, \\
\label{smooth_quench_mq}
	m_{\bq}^\text{ps} &=  \sqrt{d_\bq^2 + d_\bq^4} \left(2 n_\bq^0 +1 \right),
\end{align}
with $d_\bq\equiv (\epsilon_\bq -\epsilon_\bq^0)/(2\sqrt{\epsilon_\bq\epsilon_\bq^0})$. At this stage, it is instructive to examine the large-$q$ asymptotics of this post-quench solution: for $q\xi\gg1$, Eq. (\ref{smooth_quench_nq}) leads to $n_{\bq}^\text{ps} \simeq [m \rho_0 (g- g_0)]^2/q^4$. An instantaneous interaction quench thus turns the pre-quench exponential decay (\ref{eq:nq0_BE}) into an algebraic one, provoking a logarithmic divergence of the total energy $\int_\bq \epsilon_\bq n_{\bq}^\text{ps}$ of the system after the quench. This underlines the somewhat pathological character of the instantaneous quench for a quantum gas, which needs to be regularized by taking into account the finite duration $\tau_s$ of the interaction change. Note that a similar divergence occurs in the problem of Tan's contact in Bose gases, originating from the zero-range character of the contact interaction \cite{Tan2008,Stringari_pitaevskii2003}.  

\begin{figure}
\includegraphics[scale=0.82]{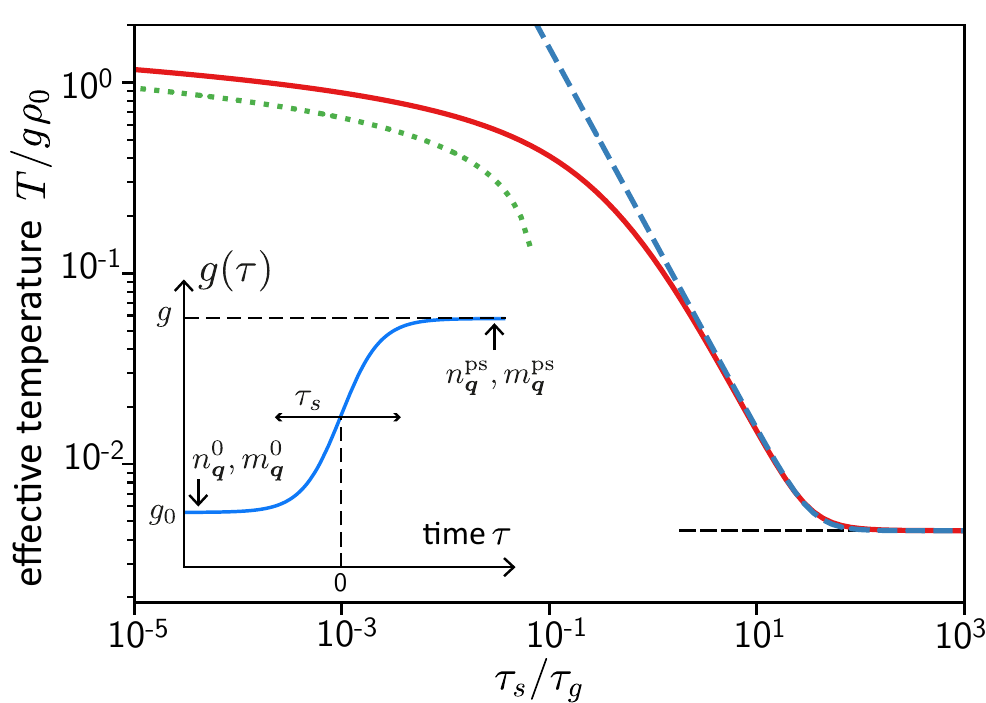}
\caption{Inset: Sketch of the Wood-Saxon function modeling an interaction quench of finite duration, with the asymptotic limits $g(-\infty)=g_0$ and $g(\infty)=g_1$. Main panel: effective equilibrium temperature reached by the Bose gas a long time after the quench as a function of the quench duration $\tau_s$ [expressed in units of $\tau_g\equiv 1/(g\rho_0)$]. 
The dotted and dashed curves show the asymptotic limits of the temperature for fast and slow quenches, Eqs. (\ref{T1_petit_taus}) and (\ref{T1_grand_taus}), respectively.
Parameters  are set to $g_0\rho_0 = 0.1, ~ g\rho_0 = 0.5, ~ T_0/g \rho_0 =0.01, ~ \rho_0\xi^2= 0.5 $. 
\label{fig_Tplot}}
\end{figure}

To overcome the ultraviolet divergence resulting from an instantaneous interaction quench, we rather consider  the smooth Wood-Saxon quench $g(\tau)=g+(g_0-g)/(1+e^{\tau/\tau_s})$, which was revisited recently in \cite{Martone2018}  and is sketched in the inset of Fig. \ref{fig_Tplot}.
For this model, Eqs. (\ref{smooth_quench_nq}) and (\ref{smooth_quench_mq}) still hold but $n_\bq^0$ ($m_\bq^0$) and $n_{\bq}^\text{ps}$ ($m_{\bq}^\text{ps}$) should now be understood as the normal (anomalous) momentum distributions a long time $|\tau|\gg\tau_s$ before and after the interaction jump, respectively, and $d_\bq^2$ is now given by \cite{Martone2018}
\begin{equation}
\label{eq:dqdef}
	d_\bq^2 = \frac{ \sinh^2\left[ \pi (\epsilon_\bq^{\phantom{0}} - \epsilon_\bq^0 ) \tau_s \right] }{\sinh (2 \pi \epsilon_\bq^0 \tau_s )  \sinh (2 \pi \epsilon_\bq^{\phantom{0}} \tau_s )}.
\end{equation}
At low momentum, the post-quench momentum distribution obeys the asymptotic law
\begin{equation}
n_{\bq}^\text{ps}\simeq \frac{T_0}{q}\frac{c^2+c_0^2}{2c c_0^2},
\end{equation}
that involves the pre-quench $c_0=\sqrt{g_0\rho_0/m}$ and post-quench $c=\sqrt{g\rho_0/m}$ speeds of sound. At large momentum, on the other hand, we have
\begin{equation}
n_{\bq}^\text{ps}\propto \exp(-2\pi\tau_s \bq^2/m).
\end{equation}
This  asymptotic law is similar to that to the pre-quench thermal distribution, $n_{\bq}^0\sim \exp[-\bq^2/(2mT_0)]$, except that the inverse of the quench duration $1/\tau_s$ now plays the role of the pre-quench equilibrium temperature.

\subsection{Momentum distributions and thermalization}

\begin{figure}[h]
\includegraphics[scale=0.9]{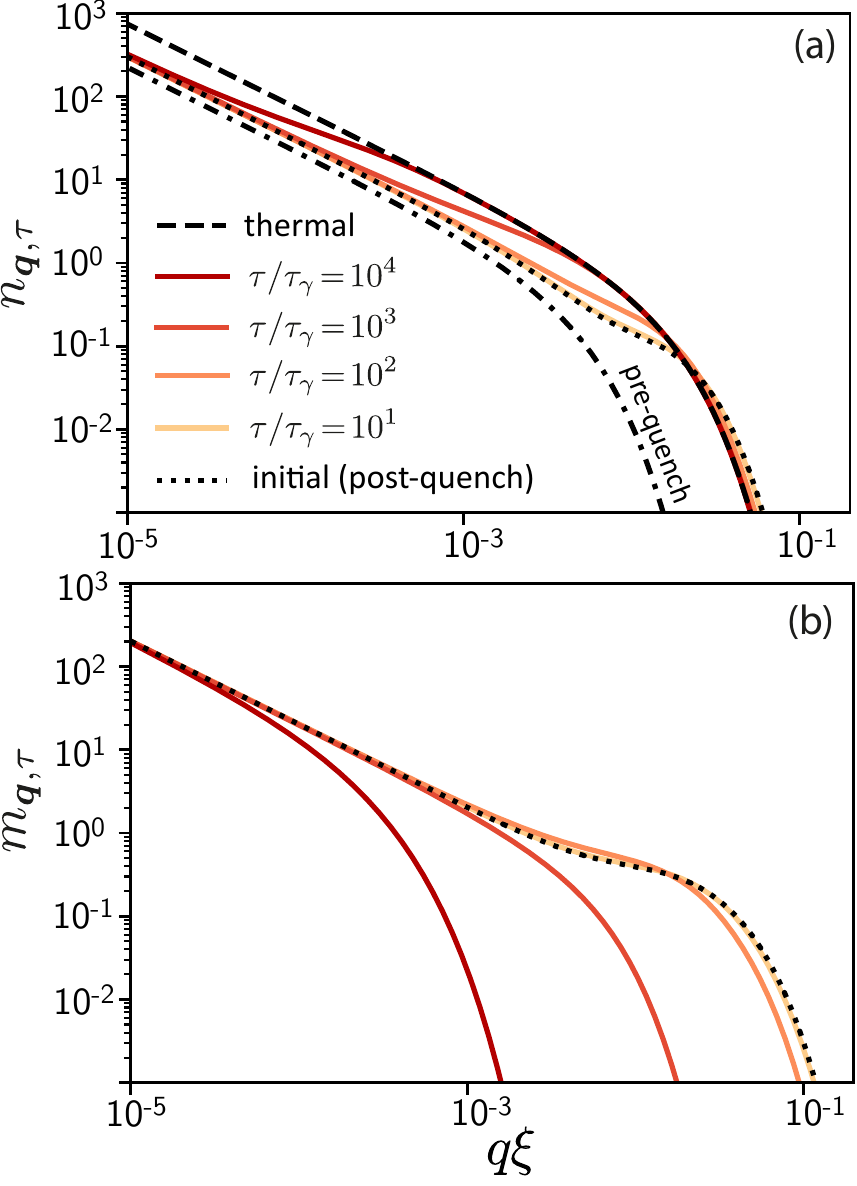}
\caption{
Time evolution of the phonon (a) normal $n_{\bq,\tau}$ and (b) anomalous $m_{\bq,\tau}$ momentum distributions following an interaction quench $g_0\to g$ near $\tau=0$. Here we set $g_0\rho_0=0.1$, $g\rho_0=0.5$, $\rho_0\xi^2=0.5$, $\tau_s/\tau_g=10$ and $T_0/g\rho_0=0.01$.
The dashed-dotted curve shows the pre-quench thermal law (\ref{eq:nq0_BE}), while dotted curves are the post-quench distributions computed from Eqs. (\ref{smooth_quench_nq}) and (\ref{smooth_quench_mq}), used as initial conditions for the kinetic equations. At long time,  $n_{\bq,\tau}$ converges to the thermal distribution (\ref{eq:nq_thermal}) (dashed curve), with an equilibrium temperature well approximated by Eq. (\ref{T1_grand_taus}).
\label{fig_nq_mq_plot}}
\end{figure}

Using the post-quench distributions (\ref{smooth_quench_nq}) and (\ref{smooth_quench_mq}) as initial conditions, we have performed numerical simulations of the kinetic equations (\ref{KE_born_nq}) and  (\ref{KE_born_mq}).
The resulting distributions $n_{\bq,\tau}$ and $m_{\bq,\tau}$ are shown in Fig. \ref{fig_nq_mq_plot} for different times.
As expected, we find that $n_{\bq,\tau}$ slowly evolves toward a thermal distribution of the form (\ref{eq:nq_thermal}) at long time. Similarly, $m_{\bq,\tau}$ converges to zero, with the region where  $m_{\bq,\tau}$ is nonzero shrinking to smaller and smaller $q-$values as time grows. For these simulations, we use as the unit time the Landau relaxation time (\ref{eq:gammaL}) evaluated at the healing length $\xi=\sqrt{1/(4 g \rho_0 m)}$:
\begin{equation}
\label{eq:taugamma}
\tau_\gamma\equiv \dfrac{1}{2\gamma^L_{q=1/\xi}}
=\frac{8}{ \sqrt{3}\pi}\rho_0\xi^2 \dfrac{g\rho_0}{T^2}.
\end{equation}
In order for the kinetic approach presented in Sec. \ref{Sec:perturbation_theory} to be valid, the separation of time scales (\ref{eq:separationtime}) should be verified, namely $\tau_\gamma$ should be large compared to the fast time scale $\tau_g\sim 1/(c|\bq|)$ that governs the coherent dynamics of the Bogoliubov phonons. Evaluated at $q=1/\xi$, the latter defines a ``nonlinear time'' that is sometimes used as a time scale in experiments:
\begin{equation}
\tau_g\sim \dfrac{\xi}{c}\sim\dfrac{1}{g\rho_0}\ll \tau_\gamma.
\end{equation}
From the definition (\ref{eq:taugamma}) of $ \tau_\gamma$, we find that in practice, this inequality  holds as long as the long-time equilibrium temperature is low enough, precisely when the product $(\rho_0\xi^2)(g\rho_0/T)^2\gg1$.

\subsection{Equilibrium temperature}

The thermal distribution (\ref{eq:nq_thermal}) reached at long time $\tau\gg\tau_\gamma$ is represented by the dashed curve in Fig. \ref{fig_nq_mq_plot}(a). The associated equilibrium temperature $T$ is entirely determined from energy conservation during the time evolution:
\begin{equation}
\int_\bq\dfrac{c|\bq|}{\exp(c|\bq|/T)-1}=\dfrac{\zeta(3)T^3}{\pi \rho_0 c^2}= \int_\bq\, \epsilon_\bq n_{\bq}^\text{ps}.
\end{equation}
The temperature $T$, computed from this relation using Eqs. (\ref{smooth_quench_nq}) and (\ref{eq:dqdef}), is displayed in Fig. \ref{fig_Tplot} as a function of the quench duration $\tau_s$. As intuition suggests, $T$ decreases when $\tau_s$ increases, i.e., as the quench is more and more adiabatic. The temperature admits a simple expression in the asymptotic regimes $\tau_s\gg\tau_g$ (slow quench) and $\tau_s\ll\tau_g$ (fast quench). For the fast quench we find
\begin{equation} \label{T1_petit_taus}
	T \sim  \sqrt{ \frac{ 3(g - g_0)^2 \rho_0^2 }{ \pi^2}  \, \log \left[ \sqrt{\frac{\tau_g}{4 \pi \tau_s}}\right]  },\quad \tau_s / \tau_g \ll 1
\end{equation}
while, for the slow quench,
\begin{equation} \label{T1_grand_taus}
	T \simeq \left[ \left(  \frac{c T_0}{c_0}\right)^3  +  \frac{\pi  (c - c_0)^2}{2^9 \zeta(3) c_0 c \tau_s^3}  \right]^{1/3}\!\!\!\!\!,\quad \tau_s / \tau_g \gg 1.
\end{equation}\\
Both Eqs. (\ref{T1_petit_taus}) and (\ref{T1_grand_taus}) are shown in Fig. \ref{fig_Tplot}, together with the exact result. The temperature is minimal for infinitely slow quenches $\tau_s/\tau_g\to\infty$, reaching  $T\to (c/c_0)T_0$. As a remark, the curve in Fig. \ref{fig_Tplot} also indicates that when $\tau_s$ is of the order of $\tau_g$ or larger, the equilibrium temperature is such that $T\ll g\rho_0$. In this limit, the quasi-particles typically belong to the phononic branch of the dispersion and, at the same time, the condition of separation of time scales is well satisfied. For this reason, in all subsequent numerical simulations we have chosen $\tau_s=10\tau_g$.

\subsection{Non-equilibrium structure factor}

To illustrate the concrete impact of the phonon relaxation dynamics in a 2D  quenched superfluid, we now study a specific observable, the non-equilibrium quantum structure factor $S_{\bq,\tau}\equiv  \langle \delta \hat{\rho}_{\bq,\tau} \delta \hat{\rho}_{-\bq,\tau} \rangle$. The structure factor is the Fourier transform of the spatial density-density correlator of the superfluid. This quantity has been recently measured experimentally, in an ultra-cold Bose gas in two dimensions \cite{Hung2013}, and in a quantum fluid of light produced in a hot atomic vapor \cite{Steinhauer2022}, both experiments involving a quench of the interaction strength. In practice, the non-equilibrium structure factor provides a simple observable to characterize the dynamical emergence of interference between quasi-particles emitted at the quench, which  manifest themselves as oscillations of $S_{\bq,\tau}$ in space and time. Such oscillations, observed in laboratory superfluids, have sparked interest because they are analogous to the famous Sakharov oscillations, a characteristic feature in the anisotropy of the cosmic microwave background radiation related to the emission of acoustic  waves in the early universe \cite{Eisenstein2008}.
Employing the Bogoliubov transformations (\ref{BGtransfo1}) and (\ref{BGtransfo2}), we can rewrite the structure factor as:
\begin{align}
\label{eq:Sq_tau}
	S_{\bq,\tau}&= 
	\frac{E_\bq}{\epsilon_{\bq}} 
	\left[2 \langle \hat{a}_{\bq,\tau}^{\dagger} \hat{a}_{\bq,\tau}^{\phantom{\dagger}} \rangle + 1 +   2 \, \text{Re} \, \langle \hat{a}_{\bq,\tau}^{\phantom{\dagger}}  \hat{a}_{-\bq,\tau}^{\phantom{\dagger}} \rangle \right]\nonumber\\
	& =  \frac{E_\bq}{\epsilon_\bq} \left[  2 n_{\bq,\tau} + 1 + 2 \cos \left( 2 \epsilon_\bq \tau \right) m_{\bq,\tau} \right],
\end{align}
where in the second equality we have introduced the normal and anomalous phonon distributions.  The structure factor primarily exhibits a fast, coherent dynamics described by the term $\propto\cos{(2\epsilon_\bq \tau})$. These oscillations stem from the interference between Bogoliubov quasiparticles with momenta $\bq$ and $-\bq$  emitted at the quench. On top these oscillations, $S_{\bq,\tau}$ is characterized by a slow relaxation dynamics due to the quasi-particle interactions that make $n_{\bq,\tau}$ and   $m_{\bq,\tau}$ slowly vary in time.
\begin{figure}
\includegraphics[scale=0.9]{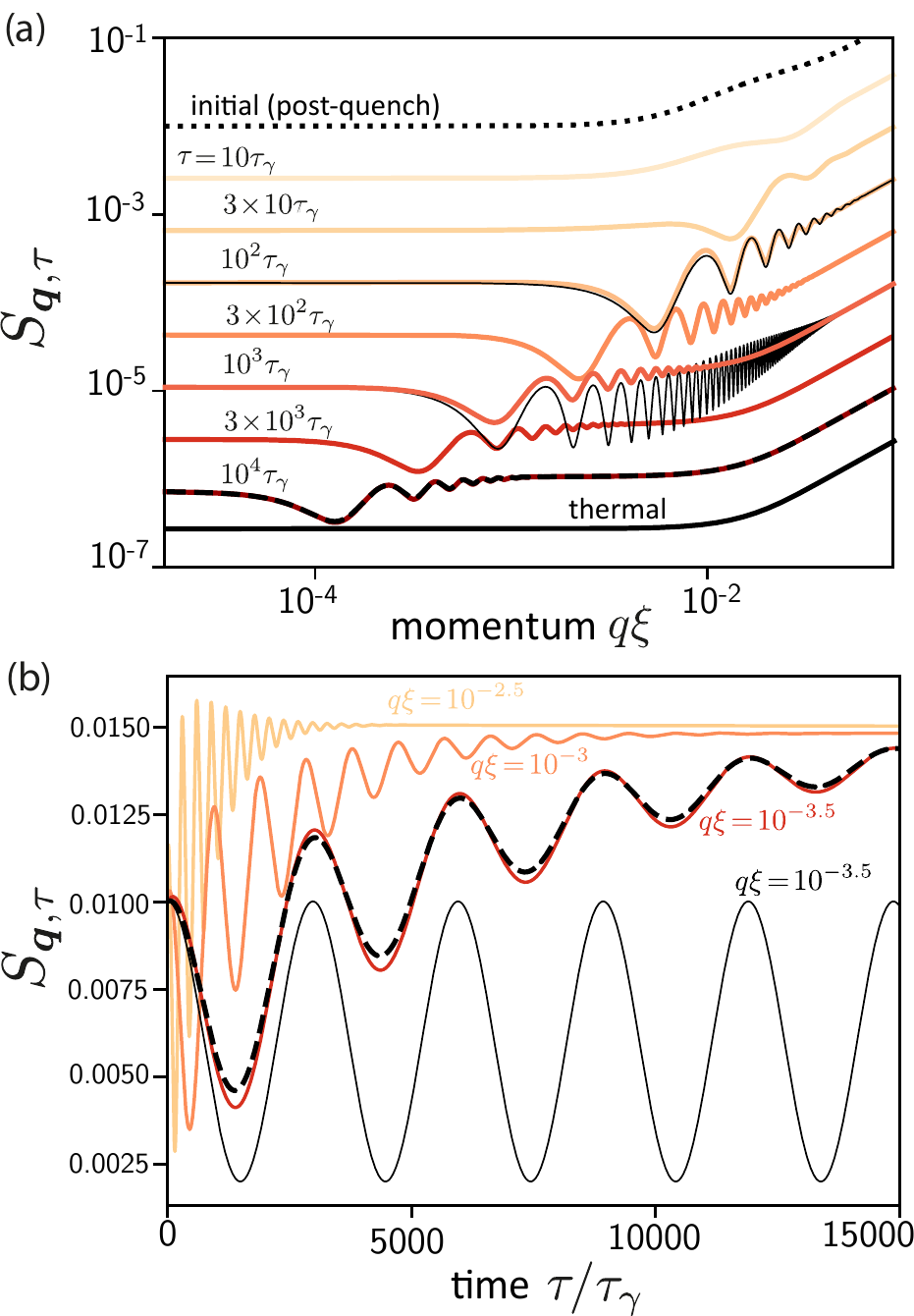}
\caption{
Non-equilibrium structure factor  $S_{\bq,\tau}$, Eq. (\ref{eq:Sq_tau}), vs. (a) momentum at different times and (b) time at different momenta. For a better readability, in panel (a) the curves are shifted downward as time increases (except $S_{\bq,\tau=0}$, black dotted curve).
In both panels, the thin black curves are the Bogoliubov prediction (\ref{eq:Sq_BG}), while the dashed curves are the long-time approximation (\ref{eq:Sq_NE}). Observe that the Bogoliubov result becomes clearly inaccurate as time increases.
Parameters have the same values as in Fig. \ref{fig_nq_mq_plot}: $g_0\rho_0=0.1$, $g\rho_0=0.5$, $\rho_0\xi^2=5 \times 10^{-4}$, $\tau_s/\tau_g=10$ and $T_0/g\rho_0=0.01$.
\label{Sq_q_plot}}
\end{figure}

The structure factor is shown in Fig. \ref{Sq_q_plot}(a) for increasing times, from its post-quench to its long-time (thermal) value. Shortly after the quench, $S_{\bq,\tau}$ exhibits sizeable oscillations of period $\pi/(c\tau)$ in momentum space. In this regime [up to $\sim 10^2\tau_\gamma$ in Fig.  \ref{Sq_q_plot}(a)], the dynamics is almost purely coherent, $m_{\bq,\tau}$ and $n_{\bq,\tau}$ remaining close to their initial, post-quench value. Within this short-time window, which was the main focus of previous experiments \cite{Hung2013, Steinhauer2022}, we can therefore approximate $m_{\bq,\tau}\simeq m_{\bq}^\text{ps}$ and $n_{\bq,\tau}\simeq n_{\bq}^\text{ps}$ in Eq. (\ref{eq:Sq_tau}), so that:
\begin{align}
\label{eq:Sq_BG}
	S_{\bq,\tau}\!\simeq\!
	 \frac{E_\bq}{\epsilon_{\bq}}
	 \text{coth}\left(\frac{\smash{\epsilon_\bq^0}}{\smash{2T_0}}\right)
	\Big[ 2d_\bq^2\!+\!1\!+\!2\sqrt{d_\bq^2\!+\!d_\bq^4}\cos(2\epsilon_\bq \tau) \Big],
\end{align}
which is nothing but the prediction of Bogoliubov perturbation theory. The approximation (\ref{eq:Sq_BG}) is shown in Fig. \ref{Sq_q_plot}(a) at both times $\tau=10^2\tau_\gamma$ and $10^3\tau_\gamma$. While in the former case  it well captures the dynamics, in the latter case it is clearly inaccurate. Indeed, at long times quasi-particle interactions become prominent and lead to a damping of the coherent oscillations. Eventually, the latter  completely disappear when the system has thermalized, with $S_{\bq,\infty}\simeq (E_\bq/\epsilon_\bq)\text{coth}(\epsilon_\bq/2T)$. While a quantitative description of $S_{\bq,\tau}$ at an arbitrary time requires a numerical resolution of the kinetic equations, at long time the phonon distributions can be approximated by their near-equilibrium expressions, obtained from Eqs. (\ref{eq:gammaL}) and (\ref{eq:gammaL_mq}). Inserting these solutions into Eq. (\ref{eq:Sq_BG}), we find
\begin{align}
\label{eq:Sq_NE}
	&S_{\bq,\tau}\simeq
	\frac{E_\bq}{\epsilon_{\bq}}
	 \text{coth}\left(\frac{\epsilon_\bq}{2T}\right)
	\Big(1- e^{-\gamma_\bq \tau} \Big)\\
	&+ \frac{E_\bq}{\epsilon_{\bq}}
	 \text{coth}\left(\frac{\smash{\epsilon_\bq^0}}{\smash{2T_0}}\right)
	\Big[ 2d_\bq^2\!+\!1\!+\!2\sqrt{d_\bq^2\!+\!d_\bq^4}\cos(2\epsilon_\bq t) \Big]e^{-\gamma_\bq \tau}\nonumber
\end{align}
where $\gamma_\bq$ coincides with either the Beliaev (\ref{eq:gammaB}) or Landau (\ref{eq:gammaL}) scattering rates depending on the range of momenta probed. In Fig. \ref{Sq_q_plot}(a), Eq. (\ref{eq:Sq_NE})  is superimposed onto the exact result for $\tau=10^4\tau_\gamma$, using $\gamma_\bq=\gamma_{\bq}^L$. The agreement is very good in the whole range of $q$ (for the chosen parameters, we have typically $cq\ll T$, such that $\gamma^L_\bq$ is always much larger than $\smash{\gamma_\bq^B}$).

The impact of the relaxation dynamics of the phonons is seen even more dramatically in Fig. \ref{Sq_q_plot}(b), which shows the structure factor at fixed momentum as a function of time. In the absence of phonon interactions (Bogoliubov approximation), $S_{\bq,\tau}$ oscillates harmonically. Comparing with the exact behavior for $q\xi=10^{-3.5}$ shows how poor the Bogoliubov approximation becomes as time grows. 

\subsection{Non-equilibrium coherence function}

As a second illustration, we study the time evolution of the coherence function of the Bose gas consecutive to an interaction quench $g_0\to g$:
\begin{equation}
G_1(r,\tau)\equiv\langle \hat{\psi}^\dagger(0,\tau)\hat{\psi}(\br,\tau)\rangle.
\end{equation}
In the density-phase representation (\ref{hydro_qu}), the coherence function can be expressed in terms of the variance of phase and density fluctuations \cite{Mora2003}:
\begin{align}
\label{G1def}
G_1(r,\tau)=&\rho_0\exp\Big\{-\frac{1}{2}\langle:[\hat{\theta}(0,\tau)-\hat{\theta}(\br,\tau)]^2:\rangle\nonumber\\
&-\frac{1}{8\rho_0}\langle:[\delta\hat{\rho}(0,\tau)-\delta\hat{\rho}(\br,\tau)]^2:\rangle\Big\},
\end{align}
where the  $:$ symbol refers to normal ordering of operators in position representation. Notice that $G_1$ only depends $r=|\br|$ due to rotation invariance. In 2D Bose gases, the spatial dependence of this function is typically dominated by the spatial growth of phase fluctuations, eventually leading to an algebraic decay of $G_1$. This behavior is noticeably different from the one of 3D Bose gases, whose phase fluctuations are very small at low temperature. 
Using the Bogoliubov transformations (\ref{BGtransfo1}-\ref{BGtransfo2}) and definitions (\ref{eq:nqdef}-\ref{eq:mqdef}), we find that Eq. (\ref{G1def})  can be rewritten as
\begin{figure}
\includegraphics[scale=0.9]{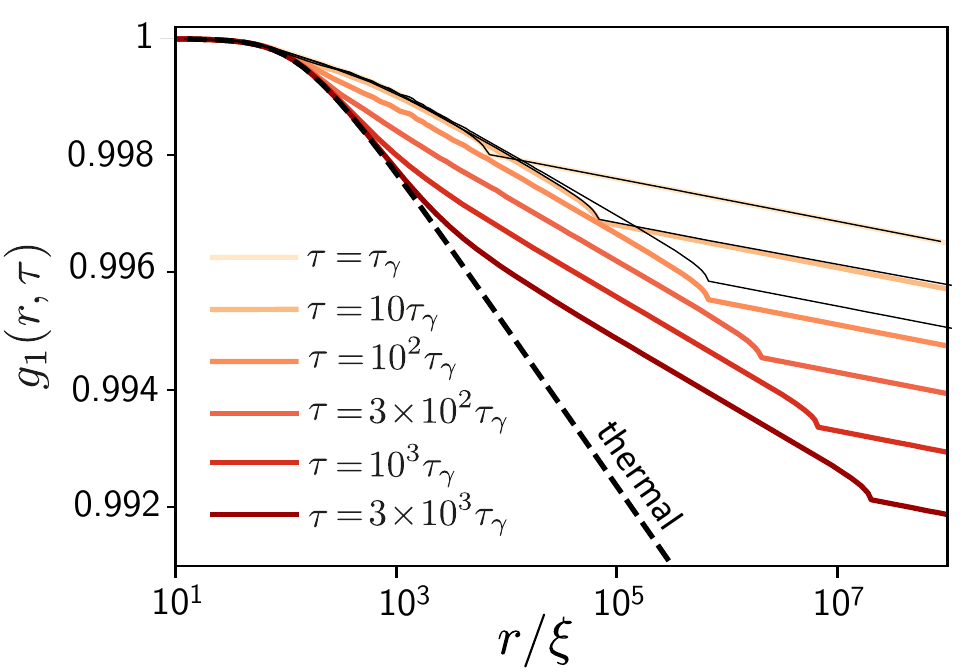}
\caption{
Non-equilibrium coherence function $g_1(r,\tau)$ vs. position $r$ at different times, computed from Eq. (\ref{eq:g1}) together with the numerical solution of quantum kinetic equations for $n_{\bq,\tau}$ and $m_{\bq,\tau}$.
The three thin black curves are the Bogoliubov prediction  at times $\tau=\tau_\gamma, 10\tau_\gamma$ and $100\tau_\gamma$. Observe that at $\tau=100\tau_\gamma$ the Bogoliubov prediction is no longer accurate. The dashed curve shows the long-time, thermal asymptotic value. Parameters are set to $g_0\rho_0=0.1$, $g\rho_0=0.5$, $\rho_0\xi^2=0.5$, $\tau_s/\tau_g=10$ and $T_0/g\rho_0=0.01$.
\label{fig:g1plot}}
\end{figure}
\begin{align}
	G_1 (r, \tau)/\rho_0\! =\! \mathcal{G}_1(r)g_1(r,\tau).
\end{align}
Here 
\begin{align}
\label{eq:g1}
 g_1(r,\tau)\!=&\exp\! \Big\{\! \! -\!   \int_\bq\frac{1}{2} 
 \left[1 \!-\! \cos (\bq\!\cdot\!\Delta \br ) \right] \Big[ \left( \frac{\epsilon_{\bq} }{E_\bq}\! +\! \frac{E_\bq }{\epsilon_\bq } \right) n_{\bq,\tau}
\nonumber \\ & 
 +\left(  \frac{E_\bq}{\epsilon_{\bq} } - \frac{ \epsilon_{\bq}  }{ E_\bq}    \right)   m_{\bq,\tau} \cos(2 \epsilon_{\bq} \tau) \Big] \Big\}
\end{align}
encodes the time evolution of the coherence following the quench.
The function $\mathcal{G}_1(r)$ is, in contrast, independent of time. It satisfies $\mathcal{G}_1(0)=1$ and quickly decays to $\mathcal{G}_1(r \gg \xi)\simeq 1-1/(16\pi\rho_0\xi^2)$ at distances larger than the healing length, a value that coincides with the quantum depletion of zero-temperature Bose gases in two dimensions. Note that $\mathcal{G}_1$ purely originates from the non-commutation of the Bogoliubov operators involved in Eq. (\ref{G1def}) and, as such, would be absent within a classical-field description. 

From now on we focus our attention on $g_1(r,\tau)$, which we have computed from Eq. (\ref{eq:g1}), using the numerical solutions of the quantum kinetic equations (\ref{KE_born_nq}) and (\ref{KE_born_mq}) for $n_{\bq,\tau}$ and $m_{\bq,\tau}$. The full time evolution of this function is shown in Fig. \ref{fig:g1plot} for $g>g_0$, and reveals  the successive emergence of three  characteristic regimes of algebraic decay. At very short times, first,  $g_1$ mainly exhibits the algebraic decay of the pre-quench equilibrium state:
\begin{equation}
g_1(r,\tau)\sim\left(\frac{\lambda_0}{r}\right)^{\frac{1}{\rho_0^{\phantom{2}}\lambda_0^2}},
\end{equation} 
with $\lambda_0=\sqrt{2\pi/(mT_0)}$ the thermal de Broglie wavelength at the (pre-quench) temperature $T_0$. Shortly after the quench, then, a second algebraic law emerges at intermediate scales, typically within a light cone of radius $r=2ct$. This characteristic decay can be described at the level of the Bogoliubov approximation, namely by simply replacing  $n_{\bq,\tau}$ and $m_{\bq,\tau}$ by their post-quench values in Eq. (\ref{eq:g1}). This leads to the ``pre-thermal'' algebraic law
\begin{equation}
g_1(r,\tau)=\left(\frac{\lambda_0}{r}\right)^{\frac{1+g/g_0}{ 2 \rho_0^{\phantom{2}}\lambda_0^2}}.
\end{equation}
At long time, finally, a third thermal algebraic scaling arises from short scales, and eventually extends to all scales as the system fully thermalizes with $n_{\bq,\tau}\to [\exp(cq/T)-1]^{-1}$ and  $m_{\bq,\tau}\to 0$:
\begin{equation}
g_1(r,\tau\to\infty)=\left(\frac{\lambda}{r}\right)^{\frac{1}{\rho_0\lambda^2}},
\end{equation}
with the algebraic exponent now controlled by the thermal wavelength $\lambda=\sqrt{2\pi/(mT)}$. Note that in the case $g>g_0$ considered here, the three algebraic exponents satisfy the inequalities
\begin{equation}
\frac{1}{\rho_0\lambda_0^2}\leq \frac{1+g/g_0}{ 2\rho_0\lambda_0^2}\leq \frac{1}{\rho_0\lambda^2},
\end{equation}
with the two bounds being inverted in the case of a down-quench $g<g_0$. It is instructive, additionally, to compare the exact shape of the coherence function with its Bogoliubov approximation. The latter is shown in Fig. \ref{fig:g1plot} for the three shortest times $\tau=\tau_\gamma, 10\tau_\gamma$ and $100\tau_\gamma$. Again, while this approximation is acceptable at short time, it becomes clearly inaccurate starting from $\tau\sim100\tau_\gamma$. This confirms that in 2D Bose gases, a description in terms of independent quasi-particles should be systematically questioned when dealing with non-equilibrium scenarios.

\section{Conclusion}
\label{Sec:conclusion}

We have presented a general theoretical framework for the many-body, non-equilibrium dynamics of 2D Bose superfluids following a quantum quench. The approach is based on a low-energy quantum hydrodynamic framework, and assumes that the time scales respectively associated with the coherent dynamics of the phonons and with the three-phonon interaction processes are well separated. Under this condition, we were able to describe the full time evolution of the phonon normal and anomalous momentum distributions, from the coherent prethermal regime to the final thermalization. As an illustration, we have applied this framework to a quantitative calculation of two commonly measured observables, the quantum structure factor and the coherence function of the superfluid following an interaction quench.

More generally, the present framework can be used to evaluate any observable, provided the latter can be expressed in terms of phonon distributions. While being a many-body quantum description, it can also be used to describe the classical-field limit, to which recent optical-fluid experiments typically belong \cite{Steinhauer2022, Abuzarli2022, Glorieux2022}. To this aim, one needs to take the limit of large occupation numbers in the kinetic equations (\ref{KE_born_nq}) and  (\ref{KE_born_mq}) and replace Bogoliubov operators by commuting scalar fields when expressing observables in terms of phonon momentum distributions. At the level of the field theory, this is done by dropping interaction processes involving quantum field variables only \cite{Berges2017}.

Being based on a low-energy framework, our approach requires the Bose gas to remain in a superfluid state at the end of the dynamical evolution. Said differently, the final equilibrium temperature should be typically below the critical Kosterlitz-Thouless temperature. While the quench dynamics of 2D Bose gases across the Kosterlitz-Thouless transition has been recently studied numerically \cite{Mathey2017, Comaron2019, Scoquart2022}, a general many-body description of this problem is, to our knowledge, still lacking. Finally, exploring the strong-interaction regime in two dimensions, where corrections to the Beliaev and Landau relaxation rates are expected \cite{Sinner2009, Sinner2010}, would be another interesting challenge for future work.


\end{document}